\documentclass[reprint, amsmath, amssymb, superscriptaddress, prl, aps]{revtex4-2}

\usepackage{xcolor}
\usepackage{caption}
\usepackage{subcaption}
\usepackage{graphicx}
\usepackage{dcolumn}
\usepackage{bm}
\usepackage{hyperref}
\usepackage{cleveref}

\makeatletter
\def\maketitle{
\@author@finish
\title@column\titleblock@produce
\suppressfloats[t]}
\makeatother

\graphicspath{ {Figures/}, {Figures_SM/} }
\newcommand{\mk}{\langle k \rangle}
\newcommand{\mr}{\langle r \rangle}

\definecolor{airforceblue}{rgb}{0.36, 0.54, 0.66}

\newcommand{\FigSM}[1]{\textcolor{blue}{#1}}
\newcommand{\FigMain}[1]{\textcolor{blue}{#1}}

\begin{document}

\title{Non-Markovian random walks characterize network robustness to nonlocal cascades}

\author{Angelo Valente}
\affiliation{Department of Mathematics, University of Trento, Via Sommarive 14, 38123 Povo (TN), Italy}
\author{Manlio De Domenico}
\email{manlio.dedomenico@unipd.it}
\affiliation{CoMuNe Lab, Department of Physics and Astronomy, University of Padua, Via F. Marzolo 8, 35131 Padua, Italy}
\author{Oriol Artime}
\email{oartime@fbk.eu}
\affiliation{CHuB Lab, Fondazione Bruno Kessler, Via Sommarive 18, 38123 Povo (TN), Italy}

\date{\today}

\begin{abstract}
Localized perturbations in a real-world network have the potential to trigger cascade failures at the whole system level, hindering its operations and functions. Standard approaches analytically tackling this problem are mostly based either on static descriptions, such as percolation, or on models where the failure evolves through first-neighbor connections, crucially failing to capture the nonlocal behavior typical of real cascades. We introduce a dynamical model that maps the failure propagation across the network to a self-avoiding random walk that, at each step, has a probability to perform nonlocal jumps toward operational systems' units. Despite the inherent non-Markovian nature of the process, we are able to characterize the critical behavior of the system out of equilibrium, as well as the stopping time distribution of the cascades. Our numerical experiments on synthetic and empirical biological and transportation networks are in excellent agreement with theoretical expectation, demonstrating the ability of our framework to quantify the vulnerability to nonlocal cascade failures of complex systems with interconnected structure.
\end{abstract}

\maketitle


\textit{Introduction.---} A wide variety of complex systems is structured in terms of nodes, representing systems' units, and links, encoding the different types of interactions among them. Yet, any trustworthy model aiming at reproducing observations and making principled predictions needs to incorporate some dynamical behavior on the network~\cite{newman2018networks, barrat2008dynamical}. In fact, understanding the interplay between structure and dynamics is still one of the major challenges in network science~\cite{barzel2013universality, kivela2014multilayer, battiston2020networks, porter2016dynamical, artime2017dynamics}. A central question concerns the robustness of a system against perturbations~\cite{gao2016universal}, since it can advance the development of powerful analytical techniques to explain and unravel rich phenomenology~\cite{dorogovtsev2008critical}, as well as it can provide a solid ground for informed interventions, e.g., disease containment based on contract tracing~\cite{ehealth2020mobile} and immunization~\cite{clusella2016immunization}, hate speech counter-measures in online environments~\cite{artime2020effectiveness} or the characterization of vulnerabilities in infrastructural networks against natural disasters~\cite{yang2017small}.

A main assumption behind the analysis of robustness is that for a system to be functional, it needs to be connected. Hence, a first quantitative proxy to assess the robustness to failures is~$s$, the normalized size of the largest connected component; thereby concepts and techniques from percolation theory become useful~\cite{stauffer2018introduction}. In percolation, a given fraction~$\phi \in [0,1]$ of nodes (or links), either selected uniformly at random or based on topological or nontopological descriptors~\cite{callaway2000network, dorogovtsev2006k, artime2021percolation}, is removed from the network. Other quantities are computed along with $s$ as a function of~$\phi$, showing interesting phenomenology such as multiple~\cite{colomer2014double,hackett2016bond} or abrupt phase transitions~\cite{achlioptas2009explosive}, and a robust-yet-fragile effect for networks with a broad enough degree distribution~\cite{albert2000error}. Percolation quantities are also employed to assess the robustness under cascading failure problems, such as the value of~$s$ when the cascade stops spreading. These models are dynamical, in the sense that a small perturbation placed in the network evolves according to some rules~\cite{watts2002simple, brummitt2012suppressing, yu2016system, huang2013cascading, shekhtman2016recent}, which depend on the phenomenon one is trying to model. For the sake of mathematical tractability, cascades are assumed to spread via direct contacts.

Be it because the physical mechanisms behind the failure propagation permit far-off malfunctions, be it because the knowledge on the observed network topology is incomplete and the failure propagates through hidden or unobserved edges, real-world cascades display nonlocal features. To name but a few examples of empirical nonlocal cascades, where node or links failures did not always occur in the neighborhood of previous ones, we have the 1996 disturbance of the Western Systems Coordinating Council (WSCC) system~\cite{NERC2002system}, the 2003 blackout in the northeastern region of the USA~\cite{nerc2004technical}, or the air-traffic disruption due to the eruption of the Icelandic volcano Eyjafjallajökul~\cite{eurocontrol2010ashcloud, wikipedia}. From a modeling standpoint, some mechanisms like flow redistribution can lead to nonlocal spreading of failures~\cite{motter2002cascade, crucitti2004model, zhang2018cascading}---with the possibility of abrupt transitions, see e.g., Refs.~\cite{zhang2018cascading, kornbluth2018network, artime2020abrupt}--- but the mathematical treatment has been hitherto under-researched due to its sophistication and there is no direct way to control the underlying properties of the nonlocal events, seriously undermining our understanding of the phenomenon.

In this article we overcome these longstanding limitations by introducing an analytically solvable model based on a class of self-avoiding random walks (SARW)~\cite{madras2013self}, which are used to model failures that propagate across the network while combining, probabilistically, local transitions and nonlocal jumps. We first describe the model and then show how to compute the time-dependent degree distribution in the surviving network. We obtain important time-dependent percolation quantities and first-stop properties of the cascades which characterize the critical behavior of the process, therefore offering an estimate of the robustness of the system as a function of time, whose validity is tested in several scenarios. Finally, we validate our theory against synthetic and empirical networks.

\begin{figure}
	\centering
    \noindent\includegraphics[width=0.8 \linewidth]{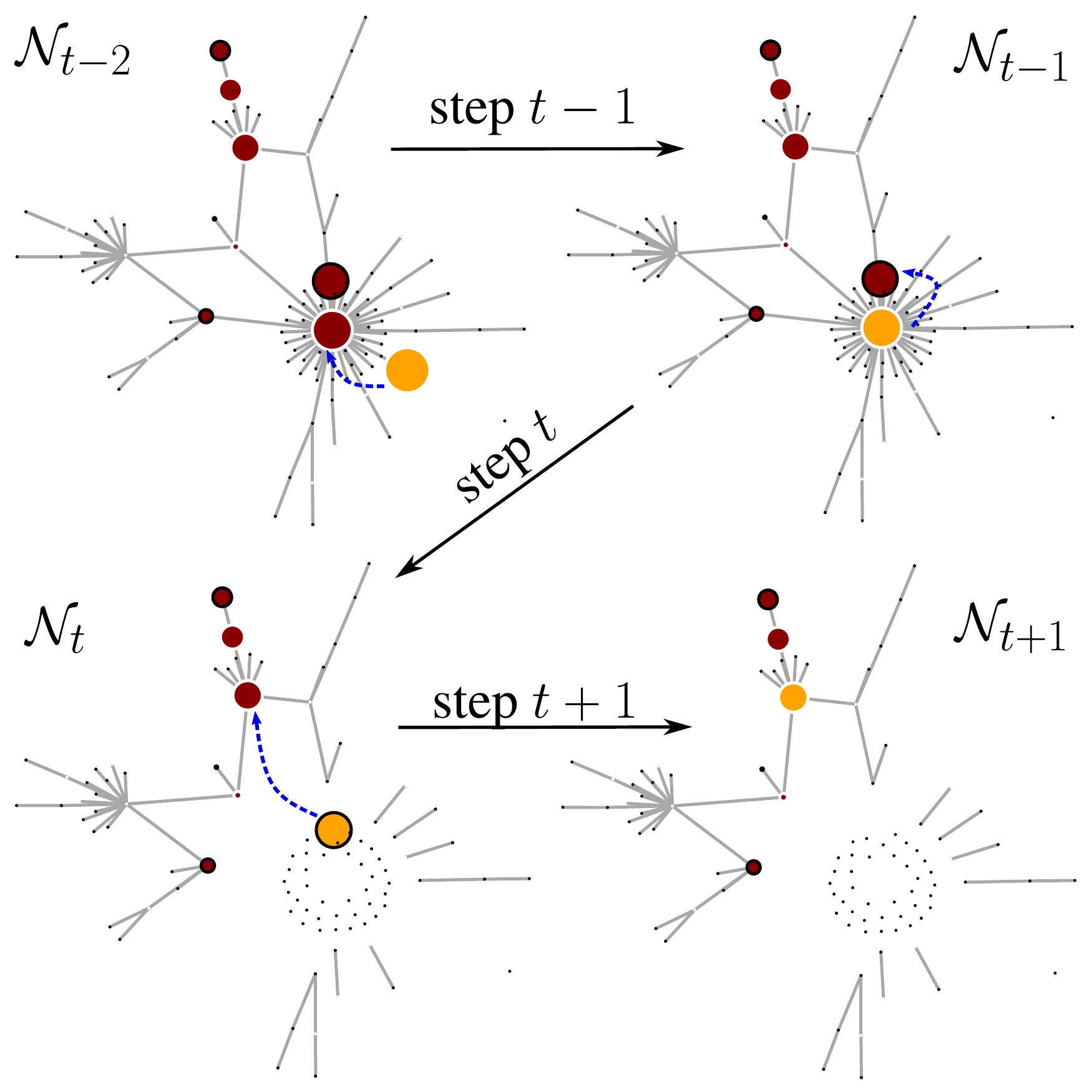}
	\caption[Excerpt of the sequence of residual networks $\{\mathcal{N}_t\}_{t \ge 0}$]{
		\label{fig:Fig1}
		Evolution of a network $\mathcal{N}_0$ under a self-avoiding teleporting random walk-- (SATRW) based dismantling process. $\mathcal{N}_{t}$ is the residual network after $t$ steps of SATRW and ``step $t$'' denotes the choice on $\mathcal{N}_{t-1}$ of the $t^{\text{th}}$ node to visit (in orange).}
\end{figure}

\textit{Mapping cascade failure to self-avoiding teleporting random walks}.--- In the following, we assume that the cascade unfolds in a timescale much faster than the recovery of nodes, and that a disrupted unit cannot be visited more than once by the failure. This fact causes the failure to be no longer Markovian and, for modeling purposes, a natural choice is to consider a SARW-like dynamics on the network. To cope with the nonlocality, we introduce a teleporting probability: At each step $t$ the failure proceeds as in a SARW ---uniformly choosing an operational neighbor and transitioning there--- with probability~$1-\alpha \in [0,1]$; otherwise, with probability~$\alpha$ it teleports to any operative node according to a teleporting rule~$T_t(k)$, in principle time- and degree-dependent; see a sketch in Fig.~\ref{fig:Fig1}. If the failure arrives at a degree-$0$ node, then it automatically teleports to an operative node. We name this stochastic process the \textit{self-avoiding teleporting random walk} (SATRW). Notice that our model interpolates between percolation, which can be seen as a purely nonlocal phenomenon, if~$\alpha=1$ and~$\phi = 1 - t/N_0$, where~$N_0 \gg 1$ is the initial network size, and the purely local process of a growing SARW when~$\alpha = 0$~\cite{herrero2005kinetic, lopez2012model, wang2019self}.

The evolution of the SATRW can be viewed as if every visited node is removed from the network, along with all the edges connected to it. This suggests to associate a time-dependent sequence of residual networks~$\{\mathcal{N}_t\}_{t\ge 0}$ to each possible SATRW on a fixed graph~$\mathcal{N}_0$ and study its average evolution. Narrowing up to configuration model networks, the natural way to characterize such a sequence is via the temporal degree distributions~$\{p_t\}_{t\ge 0}$. Similarly, we can define the time-dependent excess distribution~$\{q_t\}_{t\ge 0}$~\cite{newman2001random} and a new quantity~$\{d_t\}_{t\ge 1}$, convenient for the mathematical treatment, standing for the time-dependent probability distribution governing the degree of the node visited at step~$t$, namely the transition from $\mathcal{N}_{t-1}$ to $\mathcal{N}_t$. For example, $d_t(0)$ denotes the probability that the walker arrives at a degree-$0$ node in $\mathcal{N}_{t-1}$, whose deletion leads to $\mathcal{N}_t$. Two teleportation rules are considered. If the probability to teleport to a degree-$r$ node on~$\mathcal{N}_s$ is denoted as~$T_s(r)$, then the \textit{uniform teleportation} is
\begin{equation}
	T_{t}(r) = p_{t}(r),
\end{equation}
where a node is chosen uniformly at random in the residual network, and the \textit{biased teleportation} is 
\begin{align}
	T_{t}(r) &= 
	\frac{r \, p_{t}(r)}{\sum_s s \, p_{t}(s)} = \frac{r \, p_{t}(r)}{\mk_{t}} \notag \\
	&= q_{t}(r-1),
\end{align}
where a node is chosen with probability proportional to its degree in the residual network, see the Supplementary Material (SM) for details. By definition,~$q_{s}(-1):=0$. 


\textit{Degree distributions and giant components.---} It is possible to recursively express~$p_t$ and~$d_t$ in terms of variables in times $t-1$ and $t-2$, resulting in an effective quasi-Markovianity despite the infinite memory of the self-avoiding behavior. We obtain (see the Appendix for details)
\begin{multline}
	\label{eq: p_t}
	p_t(k) \approx \frac{1}{N_t} \bigg[
	p_{t-1}(k)N_{t-1} - d_t(k) \;+ \\ +\mr_t \big[q_{t-1}(k)-q_{t-1}(k-1)\big]
	\bigg],
\end{multline}
valid for degrees~$k = 0,1,\ldots,N_{t}-1$, where~$N_s:=N-s$ is the number of nodes in~$\mathcal{N}_{s}$ and~$\mr_t$ is the expected value of~$d_t$. For~$r = 0,1,\ldots, N_t$, we have
\begin{multline}
    \label{eq: d_t}
	d_t(r) = d_{t-1}(0) \, T_{t-1}(r) \;+ \\
	    +(1-d_{t-1}(0)) \, \bigg(\alpha T_{t-1}(r) + (1-\alpha) q_{t-2}(r)\bigg).
\end{multline}
Solving the above system of coupled equations, we gain information about the degree distribution of the residual network and the degree-dependent probability to find the walker in a functional node. In \FigSM{Figs.~SM1--SM4} we compare the analytical approximation against simulations, finding a perfect agreement. 

With~$p_t$ at hand, we can compute the fractional size of the giant component~$s_t$ of~$\mathcal{N}_t$ through the system~\cite{newman2018networks}
\begin{equation}
	\label{eq:system_s_t}
	\begin{cases}
		s_t = 1 - g_t (u_t) \\
		u_t = h_t (u_t)
	\end{cases},
\end{equation}
where~$g_t$ and~$h_t$ are the time-dependent probability generating functions respectively of the degree and the excess degree distribution,
\begin{align}
	\label{eq: time-dependent gen functions}
	g_t (z) := \sum_{k=0}^{N_t-1} p_t(k) z^k, \quad
	h_t (z) := \sum_{k=0}^{N_t-2} q_t(k) z^k.
\end{align}

\textit{Application to synthetic networks.---} We next illustrate the validity of our theory against simulations of the SATRW nonlocal process on synthetic networks with homogeneous and heterogeneous connectivity distribution, namely, Erd\H{o}s-R\'enyi (ER) and scale-free (SF) networks, respectively defined by
\begin{eqnarray}
	p^{\text{(ER)}}(k) &\approx& e^{-\mk} \frac{\mk^k}{k!} \qquad k=0,1,\ldots,N_0-1; \\
	p^{\text{(SF)}}(k) &\propto& k^{-\gamma} \qquad k=k_{\text{min}}, \ldots, k_{\text{max}},
\end{eqnarray}
where~$\mk$ denotes the average degree and~$\gamma>0$. We add constraints to these parameters in order to generate connected networks without topological correlations, namely, $\mk \ge \log N_0$ for ER~\cite{erdos1960evolution} and~$k_{\text{min}} \ge 2$ and~$k_{\text{max}} \approx \sqrt{N_0}$ for SF~\cite{catanzaro2005generation}.

\begin{figure}
	\centering
    \noindent\includegraphics[width=1 \linewidth]{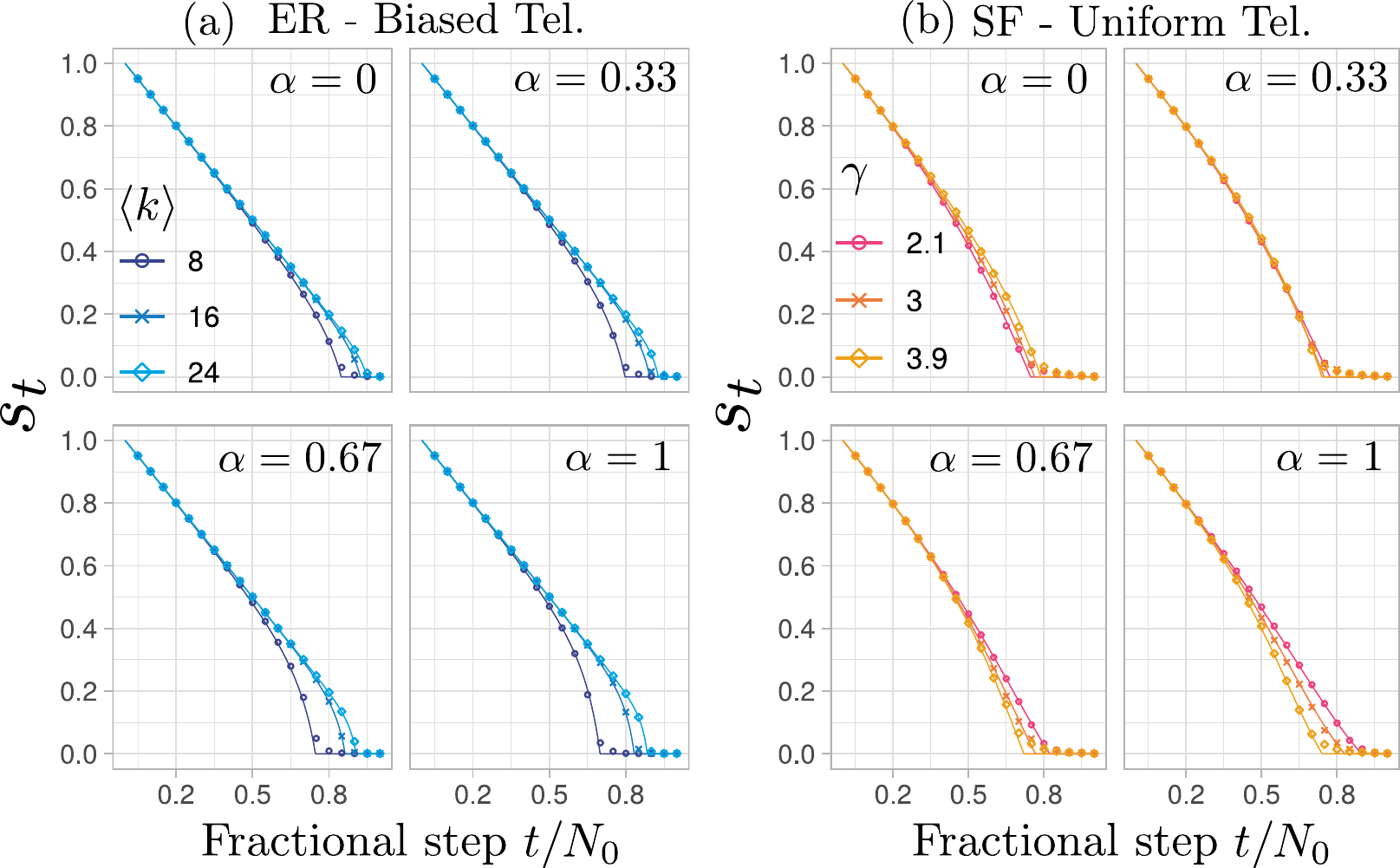}
	\caption{
		\label{fig:Fig2}%
		Evolution of the size of the giant component as a function of the fractional time~$t/N_0$. Network and teleportation rule are indicated on top (see \FigSM{Fig.~SM5} for other combinations). Each panel corresponds to a different value of teleportation parameter $\alpha$. Solid lines are theoretical predictions; circles are simulations. Initial size is~$N_0=10^3$ and averages are computed over~$25$ realizations. For SF nets, we take $k_{\text{min}} = 3$.
	}
\end{figure}

We present in Fig.~\ref{fig:Fig2} the time evolution of the size of the giant component for the different topologies and teleporting rules. A general trend that appears in all cases is that the network properties ($\langle k \rangle$ for ER, $\gamma$ for the SF; note, though, that a change in $\gamma$ induces a variation in the mean degree) and the teleportation rule do not affect the decay rate of $s_t$ at the beginning of the cascade spreading. However, there is a strong impact when approaching the dismantling point: For a fixed teleportation parameter $\alpha$, reducing the mean connectivity in ER graphs leads a faster disintegration as it could be expected. Moreover, when $\alpha$ grows, so nonlocality is enhanced, the final fragmentation occurs faster. This behavior no longer holds in SF nets. In fact, we observe that the critical point might increase or decrease when the teleportation parameter $\alpha$ is varied, evincing the nontrivial interplay that topology and nonlocality have in the robustness of interconnected systems.


\textit{Nonlocal cascades.---}So far we have assumed that when the walker steps into a degree-$0$ node, it is forced to teleport according to $T_t$ so it keeps exploring the network until all nodes are removed. However, real cascading processes normally do not dismantle the entire network but cease at a certain point leaving a part of the structure unaffected. This motivates us to incorporate a stopping criterion to the SATRW: When the walker reaches a degree-$0$ node, either it teleports with probability~$\alpha$ or it stops with probability~$1-\alpha$. We call this a \textit{nonlocal cascade}. 

To assess the robustness to nonlocal cascades, we are interested in the size of the giant component when the cascade stops,~$S^{(\text{STOP})}$. This is a stochastic variable, and we can compute its moments. To do so, first we need the stopping time distribution~$e(t)$ for~$t=1,\ldots,N_0$, that reads (see Appendix for details)
\begin{equation}
	\label{eq:e_t}
	e(t) = (1-\alpha) \, d_t(0) \prod_{i=1}^{t-1} \bigg(1-(1-\alpha)d_i(0)\bigg).
\end{equation} 

\begin{figure}
	\centering
    \includegraphics[width=1 \linewidth]{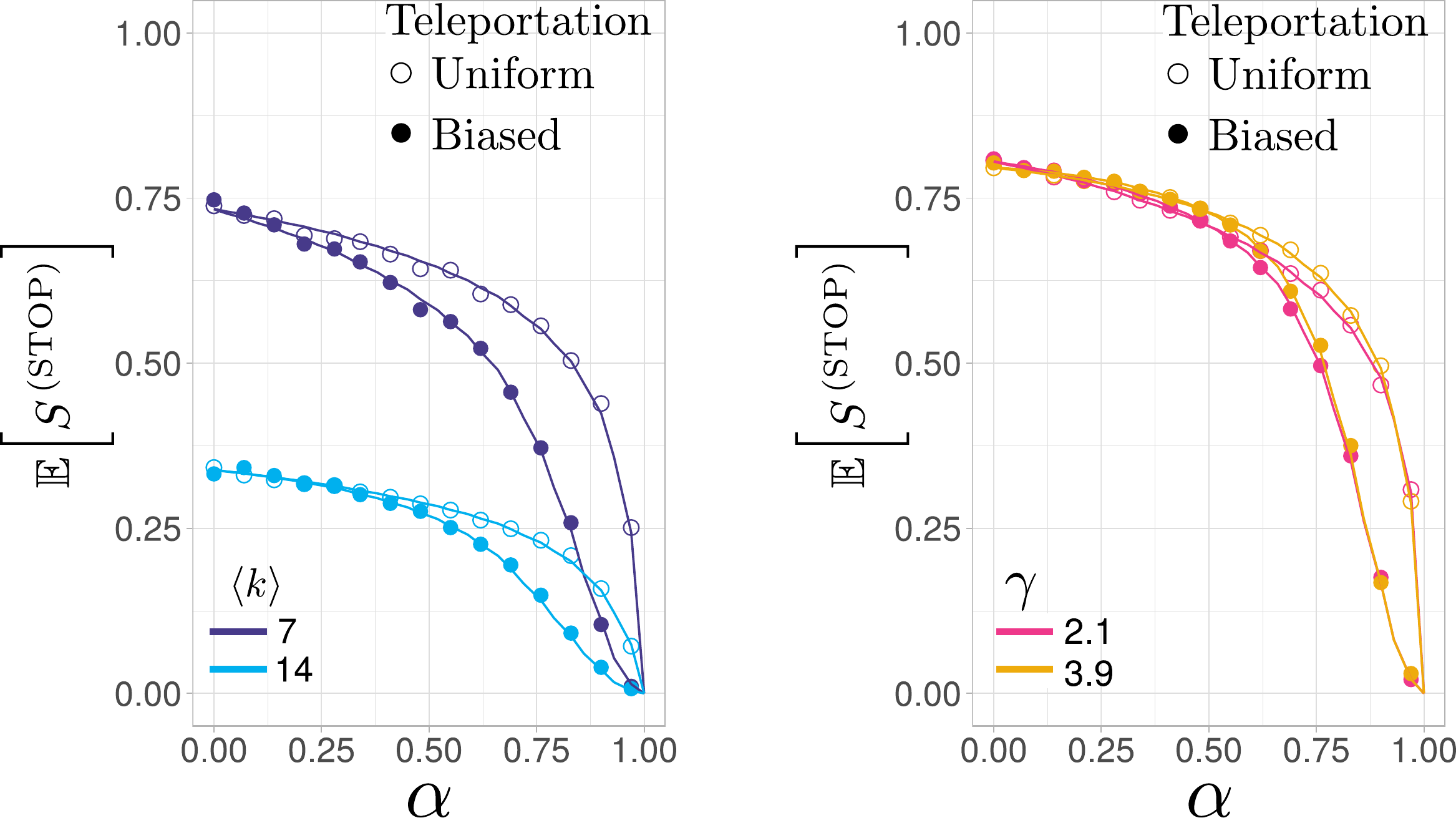} 
	\caption{
		\label{fig:Fig3}%
		Expected size of the giant component at the end of a nonlocal cascade for ER (left) and SF (right) networks. Solid lines come from theory, markers from simulations, averaged over $900$ realizations. Network size is~$N_0=10^3$, and for SF nets we set $k_{\text{min}} = 3$. See \FigSM{Fig.~SM7} for curves with other values of $\langle k \rangle$ and $\gamma$.
	}
\end{figure}

We note that an increase of the teleportation parameter is always associated with a larger stopping time, but the influence of specific network parameters strictly depends on the general topology. On one hand, in ER networks,~$e(t)$ tends to concentrate on higher times as~$\mk$ increases [see \FigSM{Fig.~SM6(a)} and \FigSM{(b)}]. On the other hand, in SF networks,~$e(t)$ is quite insensitive to~$\gamma$, and thus to the mean degree [see \FigSM{Fig.~SM6(c)} and \FigSM{(d)}]. The physical intuition behind this phenomenon is that, for fixed $\alpha$, an increase of the mean degree in ER graphs yields a uniform increase of the number of links per node, hence we expect longer stopping times as $\mk$ grows since at each time step the probability to find a functional neighbor of a previously failed node is high. For SF this is not the case, though: additional links are likely concentrated around hubs which, when disrupted, do not allow to easily find still-operational nodes in a neighborhood, leading to similar stopping times regardless the value of $\gamma$.

It is instructive to inspect the average size of the giant component at the cascade stop, given by
\begin{align}
	\label{eq:S_stop}
	\mathbb{E}\bigg[S^{\,(\text{STOP})}\bigg] = \sum_{t=1}^{N_0} s_t \, e(t),
\end{align}
a quantity that brings together the time-dependent percolation theory developed for the SATRW, Eqs.~\eqref{eq:system_s_t}, with its first-stop properties, Eq.~\eqref{eq:e_t}. We report its behavior in Fig.~\ref{fig:Fig3}. Several trends are noticeable, present in both ER and SF networks and for the two types of teleportation. First, the higher it is the nonlocality of the cascade process, the more destructive it becomes, since the probability of stopping is smaller. Second, the biased teleportation dismantles the network in a more efficient way than uniform teleportation, specially for values of the teleportation parameter $\alpha>0.5$. Third, the topology matters if $\alpha$ is not close to $1$: The hierarchical structure typical of SF networks helps to stop nonlocal cascades early, regardless of topological details governed by~$\gamma$, while ER networks have a compact and homogeneous structure and the higher the average degree, the easier it is for nonlocal cascades to spread. 

These last results seem to contradict what we reported in Fig.~\ref{fig:Fig2}: ER networks with high $\mk$ were considered robust (large critical point) under the SATRW-based dynamics, while they are found to be more fragile (low value of $S^{\,(\text{STOP})}$) to nonlocal cascades as $\mk$ increases. We interpret this as a dynamical version of the \textit{robust-yet-fragile} phenomenon: An avalanche of nonlocal failures can quickly destroy the giant component if it is able to spread, but if there is a chance for it to stop, then the topology of the network might effectively hinder such diffusion. This also holds true for SF networks, as their topology is very good at stopping cascades quite early in their evolution, but if failures are allowed to keep progressing then the network is dismantled in a similar timescale to those of ER networks, which is a topology that is not good at blocking cascades. This evinces, once more, the nontrivial relation between dynamics and topology, and sheds light on the importance of the metrics one looks at when assessing robustness.


\begin{figure}[b]
	\centering
    \includegraphics[width=0.95 \linewidth]{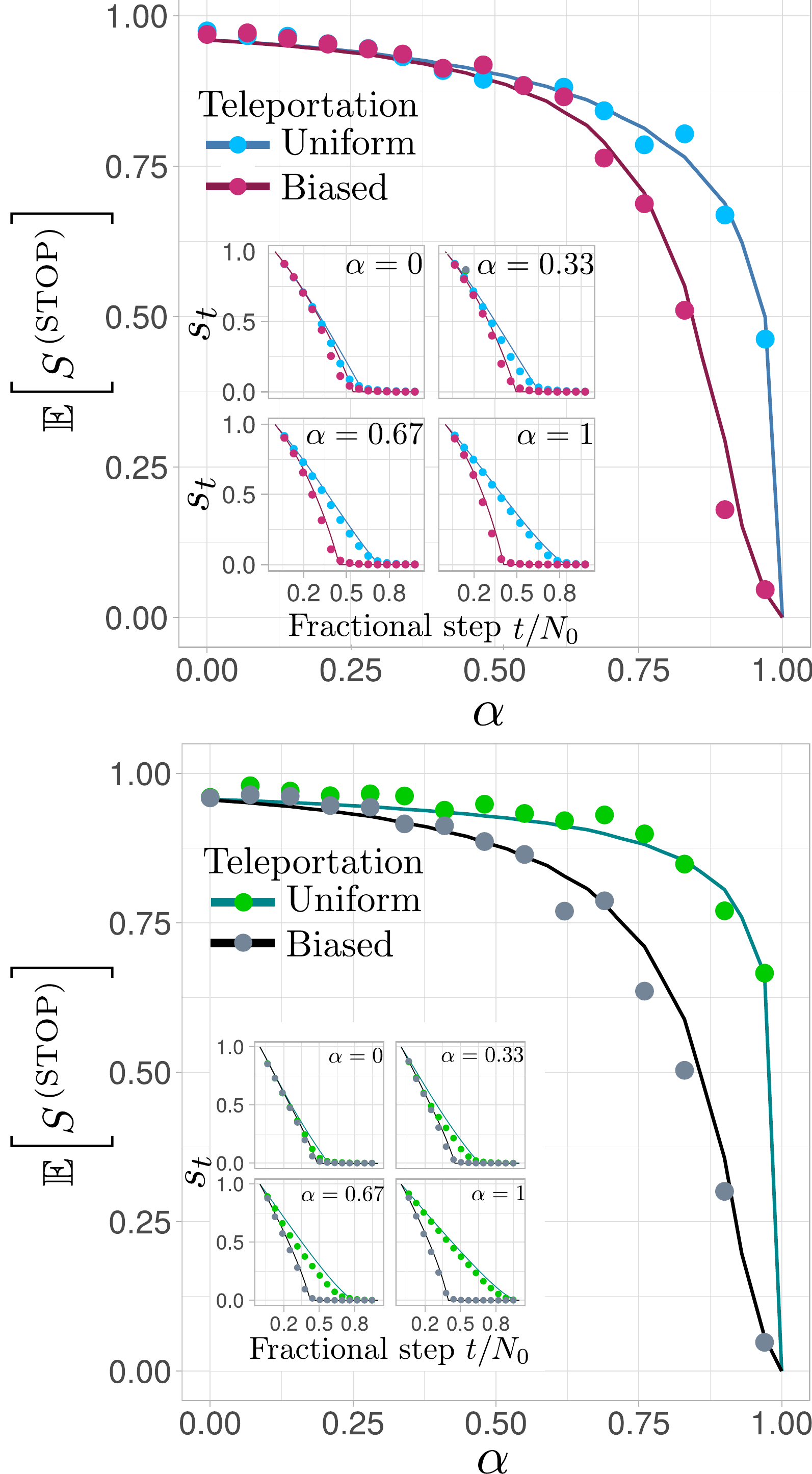}
	\caption{
		\label{fig:Fig4}
		Expected value of size of the giant component at the cascade stop as a function of the teleportation parameter $\alpha$ for the air traffic network (top) and for the \textit{C. elegans} interactome (bottom). In the insets, evolution of~$s_t$ as a function of the fractional time~$t/N_0$, for different values of $\alpha$. Solid lines come from theory, markers from simulations. Averages are over $30$ realizations. Note that multiple edges and self-loops have been discarded, directionality of links disregarded, and the resulting giant component was considered as the starting network~$\mathcal{N}_0$.
	}
\end{figure}

\textit{Application to empirical networks.---}The analytical results have been derived assuming that the degree of two adjacent nodes is not correlated and that the network is tree-like. This prompts us to ask how our theory performs when applied in empirical systems, that frequently display a wealth of different topological correlations and for which these approximations may fail. 

In the following, we show that we can capture well the evolution of the nonlocal cascade by focusing on two real topologies with a moderate value of degree assortativity~\cite{newman2018networks}. The first one is a network of air traffic routes from the Federal Aviation Administration (FAA) of the National Flight Data Center, USA~\cite{kunegis2013konect}. In this context, a node malfunction could be seen as an airport being shut down, e.g., due to meteorological events. The second network is the \textit{C. elegans} protein-protein interactome~\cite{simonis2009empirically}, where the perturbation could be understood as an initial inhibition of a certain protein, e.g., via a protein synthesis inhibitor, which is responsible for the activation/inhibition of other proteins in the PPI~\cite{ghavasieh2021multiscale}.

We show in Fig.~\ref{fig:Fig4} the results for these empirical systems. The agreement between theory and simulations is good, for both the expected value of the giant component at the cascade stop (main panel) and the SATRW dynamics without stopping criterion (insets). These results agree with the trends reported in the uncorrelated SF networks: When a significant teleporting effect is present ($\alpha$ far from zero), the targeted (biased) evolution of a failure disconnects the networks quicker than the random counterpart, but if it proceeds in cascade, the networks topology manages to stop it promptly.


\textit{Conclusions.---} Localized network disruptions in empirical settings might trigger nonlocal effects in terms of cascade failures whose propagation is usually more complex than first-neighbor search widely adopted in the literature. To better reconcile theory and observation, we have introduced a dynamical model of nonlocal failure spreading that combines local and nonlocal effects. We have characterized the rich critical behavior of our model by providing analytical expressions for several quantities employed to assess the system's robustness, such as the time-dependent degree distribution, the size of the giant component in the residual network as the process evolves, and the cascade stopping time distribution, among others. These descriptors display an excellent agreement with simulations in synthetic systems characterized by different types of complexity in terms of the heterogeneity of their structural connectivity. We find remarkable differences between homogeneous and heterogeneous systems, e.g., their dependence, or lack thereof, on the particular network parameters. However, we also report some hidden similarities between them, such as a dynamical version of the popular robust-yet-fragile feature to static attacks. It is worth noticing that, despite our framework is expected to work for locally tree-like networks lacking topological correlations, such as degree-degree ones, it still works in empirical settings as we have shown for the case of a biomolecular system, namely the interactome of the nematode \emph{C. elegans}, and an infrastructural system, namely a national air traffic network.

We envision a plethora of future generalizations for our model. One is to modify the failure dynamics itself, e.g., by including a branching mechanism of the self-avoiding walkers or by exploring more complicated stopping criteria. Similarly for the teleportation rule: it remains an open question what properties of the jumps would induce an abrupt/discontinuous percolation transition. Another direction regards the architecture on which the spreading takes place, e.g., by relaxing the assumption of uncorrelated networks and including, explicitly, topological correlations in a controlled way to understand their effect. Moreover, many networks, such as the infrastructural ones, usually show a multilayer, interdependent structure~\cite{gao2012networks, de2013mathematical, cardillo2013emergence, de2016physics}, which could be included in an adequate extension of our model. Anyway, our findings provide a solid ground for the analytical study of network robustness, in particular, and for nonlocal non-Markovian processes, in general.


\appendix
\setcounter{equation}{0}
\setcounter{figure}{0}
\renewcommand{\theequation}{A.\arabic{equation}}
\renewcommand{\thefigure}{A.\arabic{figure}}
\section{Appendix}

The effect of a Self-Avoiding Teleporting Random Walk (SATRW) on a network $\mathcal{N}_0$ is described by the time-dependent sequence of residual networks $\{\mathcal{N}_t\}_{t\ge 0}$. We are interested in studying the average evolution of such a sequence through that of two time-dependent distributions: $p_t$, which is the degree distribution of $\mathcal{N}_t$, and $d_t$, standing for the degree distribution of the node visited at step $t$ (transition from $\mathcal{N}_{t-1}$ to $\mathcal{N}_t$). In this Appendix we show how to compute both distributions, as well as the stopping time distribution when a stopping criterion is added to the evolution of the SATRW.

\begin{figure}
    \centering
    \includegraphics[width=1 \linewidth]{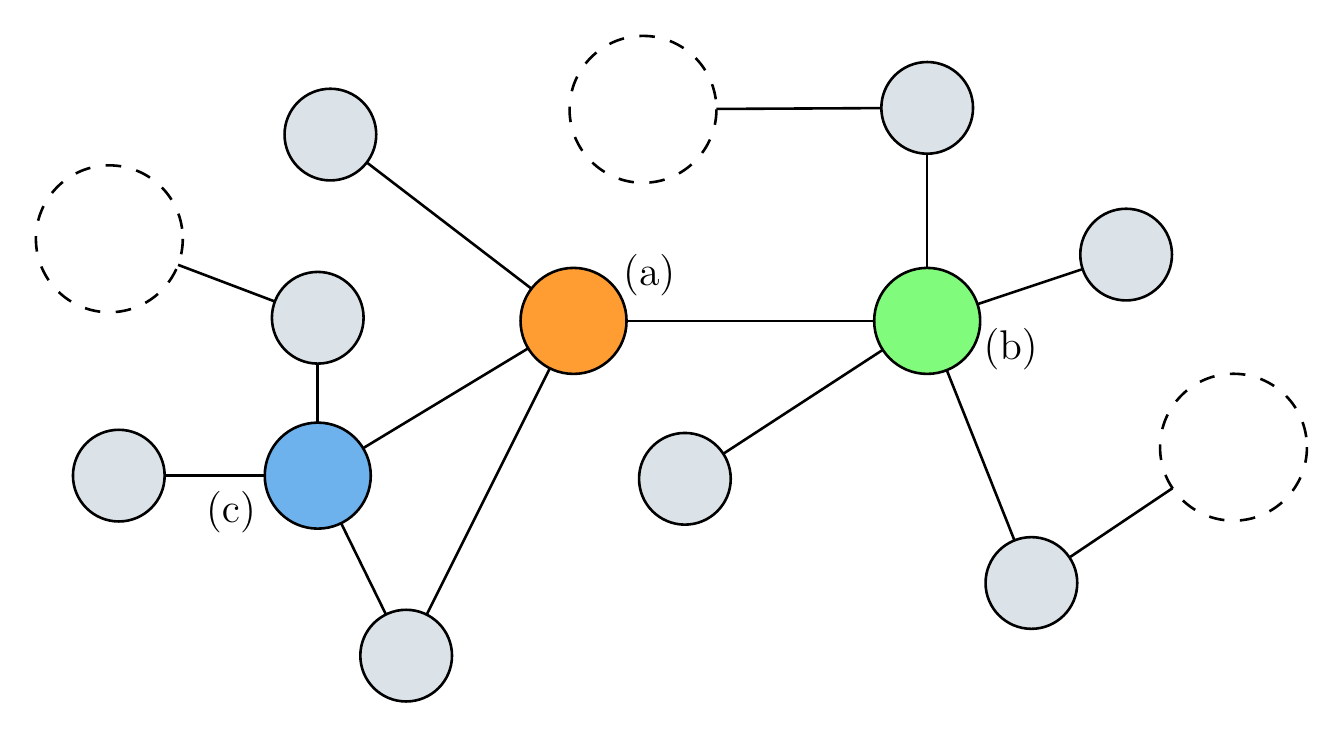}
	\caption[The three types of nodes considered in the computation of $p_t$]{
		\label{fig:FigApp}%
		The three types of nodes that cause a change from $N_{t-1}(k)$ to $N_t(k)$ are graphically represented. The picture depicts $\mathcal{N}_{t-1}$ and assumes that the orange node is visited at step $t$. The three colored nodes illustrate the cases (a), (b) and (c) discussed in the Appendix, while the big white circles represent any subgraphs that complete the network.
	}
\end{figure}

\textit{\textbf{Computation of $p_t$.}---} Rather than trying to directly derive an expression for $p_t$ we focus, for convenience, on the average number of degree-$k$ nodes in $\mathcal{N}_t$, namely
\begin{equation}
\label{eq: N_t}
N_t(k) := p_t(k) N_t.
\end{equation}
Here $N_t$ is the total number of nodes of $\mathcal{N}_t$. How does $N_t(k)$ change in one step, from $N_{t-1}(k)$ to $N_t(k)$? Analyzing the degree of the node visited at step $t$, hereinafter appealed as ``deleted", along with the degrees of its neighbours, we distinguish three different contributions to this quantity, see Fig.~\ref{fig:FigApp}:

\begin{enumerate}
	\item[(a)] The deleted node could have $k$ adjacent nodes, and this happens with probability $d_t(k)$. Thus, on average, $d_t(k)$ degree-$k$ nodes are deleted;
	
	\item[(b)] The degree-($k+1$) neighbors of the deleted node will lose an edge, then becoming degree-$k$ nodes in the next residual network $\mathcal{N}_t$. If the deleted node has degree $s$, denote the number of these neighbors in $\mathcal{N}_{t-1}$ with $N_{t-1}(k+1|s)$. Summing over all the possibilities, the number of such nodes is
	\begin{equation}
		\label{eq: degree k+1 nodes of deleted node}
		\sum_{s=0}^{N_t} d_t(s) N_{t-1}(k+1|s).
	\end{equation}
	Since configuration model networks exhibit no degree correlation, $N_{t-1}(k+1|s)$ is approximately equal to $s$ times the probability that a neighbour of a randomly chosen node has degree $k+1$ (or equivalently, excess degree $k$), leading to 
	\begin{equation}
	    N_{t-1}(k+1|s) \approx s \, q_{t-1}(k).
	\end{equation}
	The summation in Eq.~\eqref{eq: degree k+1 nodes of deleted node} then becomes
	\begin{align}
		\label{eq: degree k+1 nodes of deleted node - 2}
		\notag
		\sum_{s=0}^{N_t} d_t(s) N_{t-1}(k+1|s) &\approx
		\sum_{s=0}^{N_t} d_t(s) s \, q_{t-1}(k) \\
	    &= \mr_t \, q_{t-1}(k),
	\end{align}
	having denoted with $\mr_t$ the expected value of $d_t$.
	
	\item[(c)] The degree-$k$ neighbors of the deleted node will lose an edge too, so they will not have degree $k$ in $\mathcal{N}_t$. As before, if the deleted node has degree $s$, the number of these neighbors is 
	\begin{equation}
	    N_{t-1}(k|s) \approx s \, q_{t-1}(k-1).
	\end{equation}
	Summing over all the possibilities, the number of such nodes is
	\begin{equation}
	\sum_{s=0}^{N_t} d_t(s) N_{t-1}(k|s) \approx \mr_t \, q_{t-1}(k-1).
	\end{equation}
\end{enumerate}

Putting the pieces together, for $k=0,1,\ldots,N_t-1$
\begin{align}
&N_t(k)-N_{t-1}(k) \approx \notag \\
&\qquad\qquad- d_t(k) + \mr_t \big[q_{t-1}(k)-q_{t-1}(k-1)\big],
\end{align}
where $q_{t-1}(-1):=0$. We now come back to probabilities using Eq.~\eqref{eq: N_t}, thus obtaining
\begin{multline}
	\label{eq: p_t_app}
	p_t(k) \approx \frac{1}{N_t} \bigg[
	p_{t-1}(k)N_{t-1} - d_t(k) \;+ \\ +\mr_t \big[q_{t-1}(k)-q_{t-1}(k-1)\big]
	\bigg].
\end{multline}
Some approximations have been made to derive this expression, so the probability is not properly normalized. The normalization constant for $p_t$, $t \ge 1$, is 
\begin{align}
	C_t &= \frac{1}{N_t} \sum_{k=0}^{N_t-1} \bigg[
	p_{t-1}(k)N_{t-1} - d_t(k) \;+ \notag \\
	&\qquad\qquad +\mr_t \big[q_{t-1}(k)-q_{t-1}(k-1)\big]
	\bigg] \notag \\
	&= 1 + \frac{d_t(N_t)}{N_t} - p_{t-1}(N_t) \bigg[ 1 + \frac{1}{N_t} - \frac{\mr_t}{\mk_{t-1}} \bigg].
\end{align}
When dealing with degree distributions that assign very low probabilities to the tail, it is safe to assume $p_s(N(s)-1) \approx 0$ for any time $s \ge 0$. 
Moreover, $d_t(N_t) / N_t$ is a small number too for $t$ small and/or $N$ large, being $d_t$ a distribution.
All things considered, the normalization constant deviates from $1$ by a negligible quantity, at least when $t$ is not too large.

\textit{\textbf{Computation of $d_t$.}---} In principle, there are two ways to step on a node during the walk: the walker either reaches it by following an edge (probability $1-\alpha$) or teleports to it (probability $\alpha$).
$d_t$ is then a weighted average of two distributions with weights $1-\alpha$ and $\alpha$.

\begin{enumerate}
	\item In the teleporting case, the walker steps on a random node in $\mathcal{N}_{t-1}$ according to the specific teleportation rule. The probability to teleport to a degree-$r$ node on $\mathcal{N}_{t-1}$ is denoted as $T_{t-1}(r)$.
	
	\item In the nonteleporting case, the walker reaches a degree-$r$ node in $\mathcal{N}_{t-1}$ coming from a node present in $\mathcal{N}_{t-2}$.
	The reached node has degree $r+1$ in $\mathcal{N}_{t-2}$, therefore excess degree $r$ there. 
	Since configuration model networks exhibit no degree correlation, as we have already pointed out in the previous computations (see Eq.~\eqref{eq: degree k+1 nodes of deleted node - 2}), $q_{t-2}(r)$ is a good approximation for this event to happen, no matter the degree of the node the walker comes from.
\end{enumerate}

In practice, the degree of the previously deleted node should be considered as well because, in case this is equal to $0$, the walker must proceed via teleportation with probability $1$.
This suggests to introduce the conditional probabilities $d_t(r|0)$ and $d_t(r|0^c)$, i.e. the probability to reach a degree-$r$ node at step $t$ knowing that the walker comes from either a degree-$0$ node or not, events happening with probabilities $d_{t-1}(0)$ and $(1-d_{t-1}(0))$. In the first case we simply choose the next node via teleportation, while in the second case we weigh the two possibilities described above. 
For $r = 0,1,\ldots, N_t$:
\begin{equation}
	\label{eq: d_t_app}
	d_t(r) = d_{t-1}(0) \, d_t(r|0) + (1-d_{t-1}(0)) \, d_t(r|0^c),
\end{equation}
where the conditional probabilities are computable as
\begin{align}
	d_t(r|0) &= T_{t-1}(r); \\
	d_t(r|0^c) &= \alpha T_{t-1}(r) + (1-\alpha) q_{t-2}(r).
\end{align}
Since the first step is always a teleportation, 
\begin{equation}
	d_1(r) = T_0(r).
\end{equation}

Some comparisons between the theoretical degree distribution and the one from the simulated process are shown in \FigSM{Figs.~SM1--SM4}, depicting two particular ER and SF networks (the fractional steps for which $p_t$ reduces to a Dirac delta concentrated at $0$ are not graphically considered). We stress that theoretical predictions (solid lines) and empirical simulations (dashed lines) match so well that they are almost indistinguishable.

\textit{\textbf{Stopping time distribution.}---}For a nonlocal cascade, the stopping time $t$ is the time corresponding to the last step. Let $E_t$ denote the event \textit{``The nonlocal cascade stops at time $t$"} (equivalently, step $t$ is the last one). We denote with $\bar{e}(t)$ the conditional probability to stop at time $t$ knowing that step $t$ has been reached.
This event happens when the walker visits a degree-$0$ node at step $t$ (probability $d_t(0)$) but it is unable to proceed via teleportation (probability $1-\alpha$):
\begin{equation}
	\label{eq: e_bar(t)}
	\bar{e}(t) := \mathbb{P}[E_t | E_1^C, \ldots, E_{t-1}^C] = (1-\alpha) d_t(0).
\end{equation} 
The probability $e(t) := \mathbb{P}[E_t]$ is computable in terms of $\bar{e}(t)$. Indeed, $e(1) = \bar{e}(1)$ and for $t\ge 2$:
\begin{align}
	\notag
	e(t) &= \mathbb{P}[E_t] = 
	\mathbb{P}\left[E_t \bigg| \bigcap_{i=1}^{t-1} E_i^C\right] \mathbb{P}\left[\bigcap_{i=1}^{t-1} E_i^C\right] + \\
	\notag
	&\qquad +\mathbb{P}\left[E_t \bigg| \left( \bigcap_{i=1}^{t-1} E_i^C \right)^C \right] 
	\mathbb{P}\left[\left( \bigcap_{i=1}^{t-1} E_i^C \right)^C\right] \\ \notag
		&= \bar{e}(t) \mathbb{P}[E_1^C, \ldots, E_{t-1}^C] + 
	\mathbb{P}\left[E_t \bigg| \bigcup_{i=1}^{t-1} E_i \right] \mathbb{P}\left[\bigcup_{i=1}^{t-1} E_i \right] \\
	\label{eq: e(t)}
		&= \bar{e}(t) \, \mathbb{P}[E_1^C, \ldots, E_{t-1}^C],
\end{align}
where $\mathbb{P}[E_t | E_1 \cup \ldots \cup E_{t-1}] = 0$ since the walk cannot stop at time $t$ if it already stopped at some step $i \in \{1, \ldots, t-1\}$. The term $\mathbb{P}[E_1^C, \ldots, E_{t-1}^C]$ represents the probability to stop after step $t-1$, and it is computable by means of a recursive reasoning: denoting with 
$ec(s) := \mathbb{P}[E_1^C, \ldots, E_s^C]$, we have
\begin{align}
	\notag
	ec(s) &= \mathbb{P}[E_s^C | E_1^C, \ldots, E_{s-1}^C] \, \mathbb{P}[E_1^C, \ldots, E_{s-1}^C] \\ \notag
	&= (1 - \bar{e}(s)) \, ec(s-1) \\ \notag
	&= (1 - \bar{e}(s)) (1 - \bar{e}(s-1)) \, ec(s-2) \\ \notag
	&\enspace\vdots \\ \label{eq: ec(s)}
	&= \prod_{i=2}^{s} (1-\bar{e}(i)) \, ec(1) = \prod_{i=1}^{s} (1-\bar{e}(i)),
\end{align}
the last equality coming from 
\begin{equation}
	ec(1) = \mathbb{P}[E_1^C] = 1-\mathbb{P}[E_1] = 1-e(1) = 1-\bar{e}(1).
\end{equation}
This is coeherent with the intuition that the walk stops after a certain step if it did not stop in any of the previous ones. Putting Eqs.\eqref{eq: e_bar(t)}, \eqref{eq: e(t)} and \eqref{eq: ec(s)} together, 
\begin{align}
	\label{eq: e(t) ver2}
	\notag
	e(t) &= \bar{e}(t) \, ec(t-1) \\
	&= (1-\alpha) \, d_t(0) \prod_{i=1}^{t-1} (1-(1-\alpha)d_i(0)).
\end{align} 

In order to have a distribution with support included in the unitary interval $[0,1]$, we define the fractional stopping time distribution $\tilde{e}(s)$, defined as
\begin{equation}
	\tilde{e}(s) = e (t) \quad \text{if } s = \frac{t}{N}, \; t \in \{1, \ldots, N\}.
\end{equation}

The theoretical fractional stopping time distributions $\tilde{e}(s)$ for ER and SF networks are compared with the empirical ones in \FigSM{Fig.~SM6}. Since the duration of the cascade strictly depends on $\alpha$, we see that $\tilde{e}(s)$ peaks at higher and higher values as $\alpha$ increases. A first big difference emerging between the homogeneous and the heterogeneous case is the overall impact of the average degree on the fractional stopping time distribution, regardless of $\alpha$ and the teleportation rule. On the one hand, $\tilde{e}(s)$ shifts its mass more and more toward high times as the average degree increases in ER networks; on the other hand, $\tilde{e}(s)$ in SF networks seems not to be influenced by a change in the average degree caused by a variation of $\gamma$. 
This precludes the possibility of increasing the number of functioning (unvisited) nodes at the end of a nonlocal cascade by tuning the parameter $\gamma$ if the graph has power-law degree distribution. 

Another worth mentioning aspect is the effect of teleportation. When most of the nodes of the starting network $\mathcal{N}_0$ have already been visited, many degree-$0$ nodes will appear. The degree-biased teleportation rule will avoid the walker to teleport to such nodes, reducing the number of times the walk has the chance to stop. On the contrary, the uniform teleportation rule ignores the degrees of the residual nodes, and it is therefore more likely to hop to isolated nodes, leading to smaller stopping times. However, this difference is evident only if $\alpha$ is big enough, otherwise the walk is expected to stop after a few times the walker has found itself in degree-$0$ nodes.

\bibliography{biblio}

\newpage
$ $
\newpage

\setcounter{figure}{0}
\renewcommand{\thefigure}{SM\arabic{figure}}
\renewcommand{\theequation}{SM\arabic{equation}}

\title{Supplemental Material for \textit{Non-Markovian random walks characterize network robustness to nonlocal cascades}}

\maketitle

\onecolumngrid
In this supplementary material we provide some supporting figures to the results discussed in the main paper.

\begin{figure*}
\centering
\captionsetup{justification=raggedright}
\caption{
	\label{fig:SM1}%
	\small
	Comparison between theoretical predictions and empirical estimations of $p_t$ for several $\alpha$ on \textbf{ER} networks with $N_0=10^3$ and $\mk = 7$ in case of \textbf{uniform} teleportation.}

    \includegraphics[width=0.9 \linewidth]{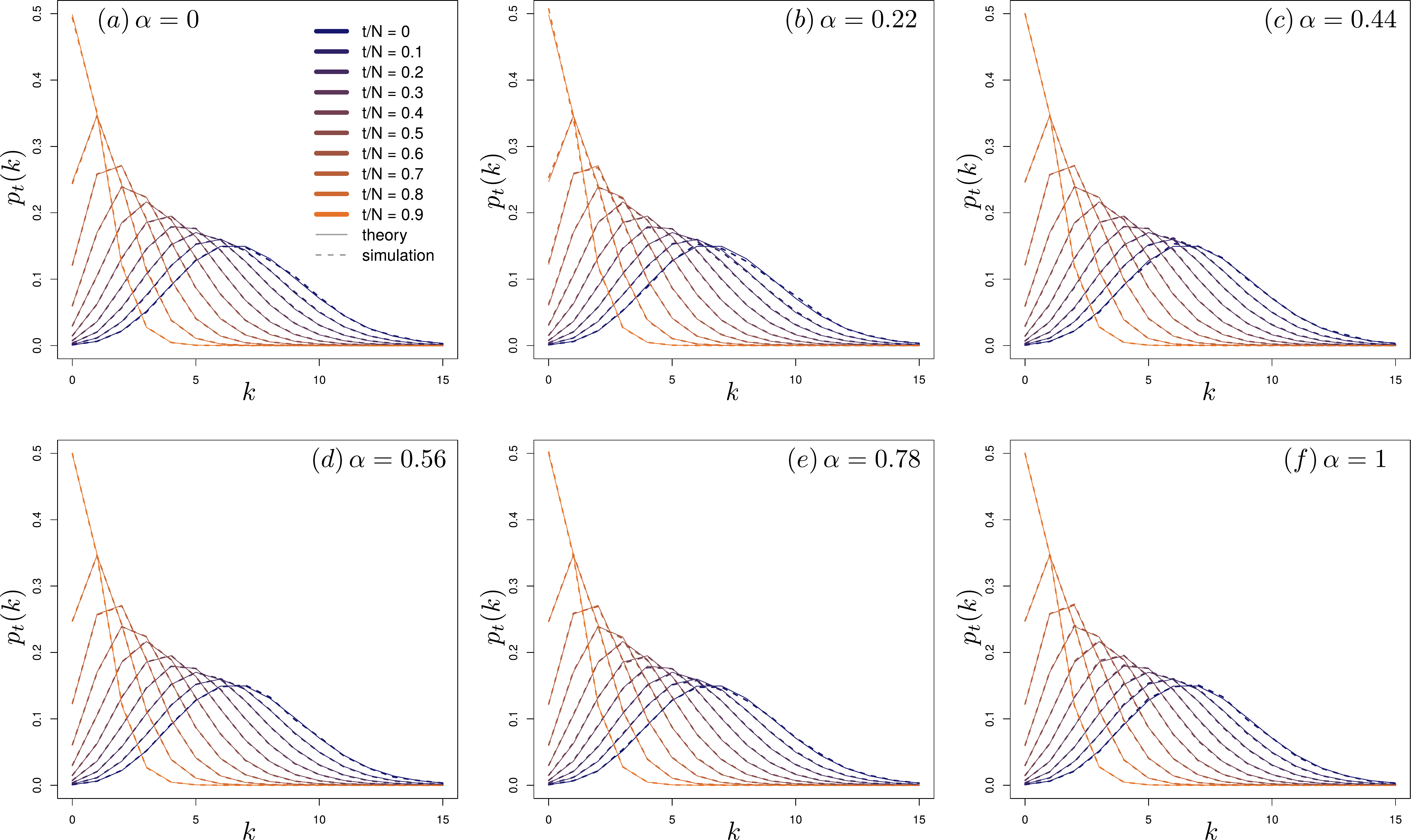}

\end{figure*}

\begin{figure*}
\centering
\captionsetup{justification=raggedright}
\caption{
	\label{fig:SM2}%
	\small
	Comparison between theoretical predictions and empirical estimations of $p_t$ for several $\alpha$ on \textbf{ER} networks with $N_0=10^3$ and $\mk = 7$ in case of \textbf{biased} teleportation.}
    
     \includegraphics[width=0.9 \linewidth]{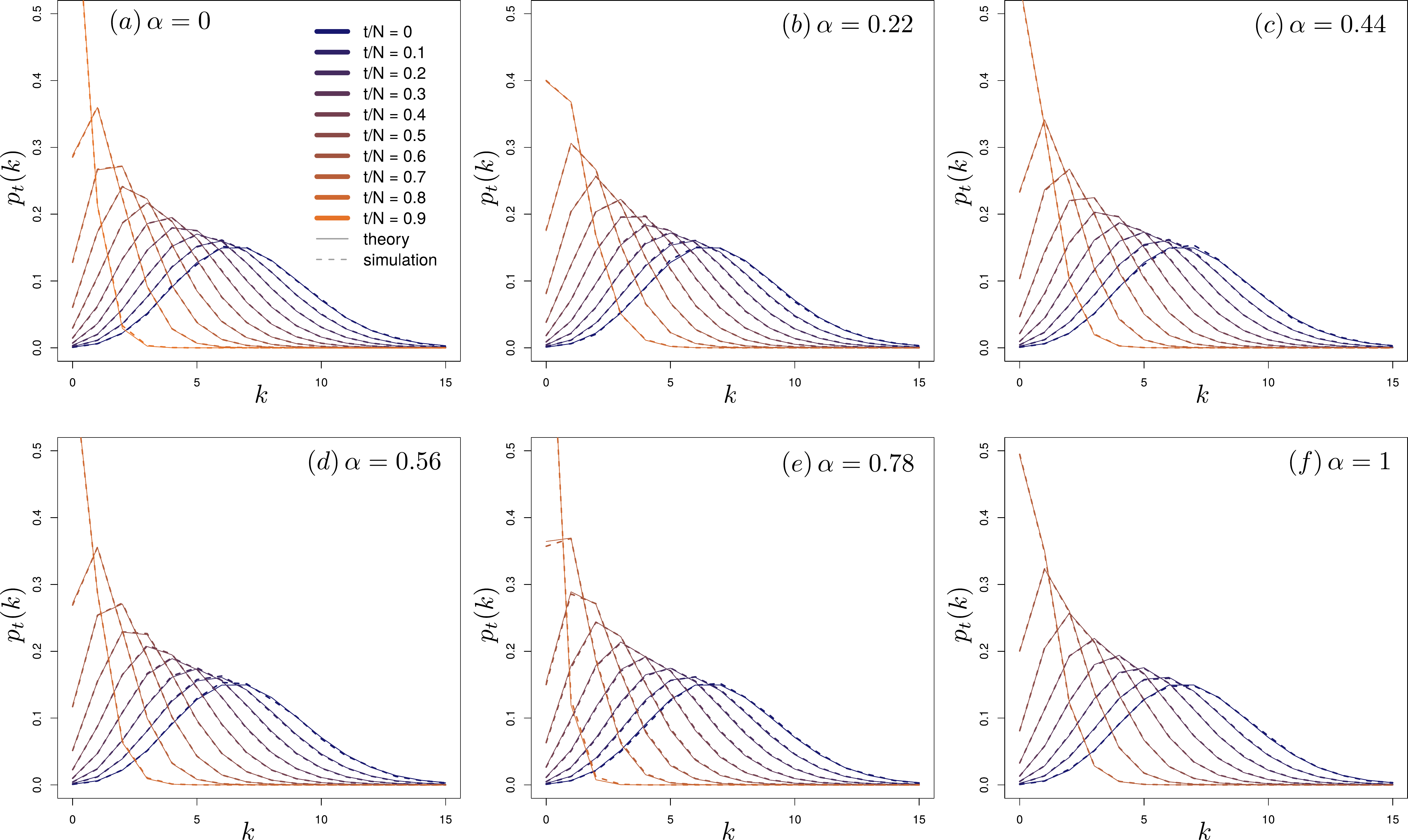}

\end{figure*}

\begin{figure*}
\centering
\captionsetup{justification=raggedright}
\caption{
	\label{fig:SM3}%
	\small
	Comparison between theoretical predictions and empirical estimations of $p_t$ for several $\alpha$ on \textbf{SF} networks with $N_0=10^3$, $\gamma = 2.5$ and $k_{\text{min}} = 3$ in case of \textbf{uniform} teleportation.}

    \includegraphics[width=0.9 \linewidth]{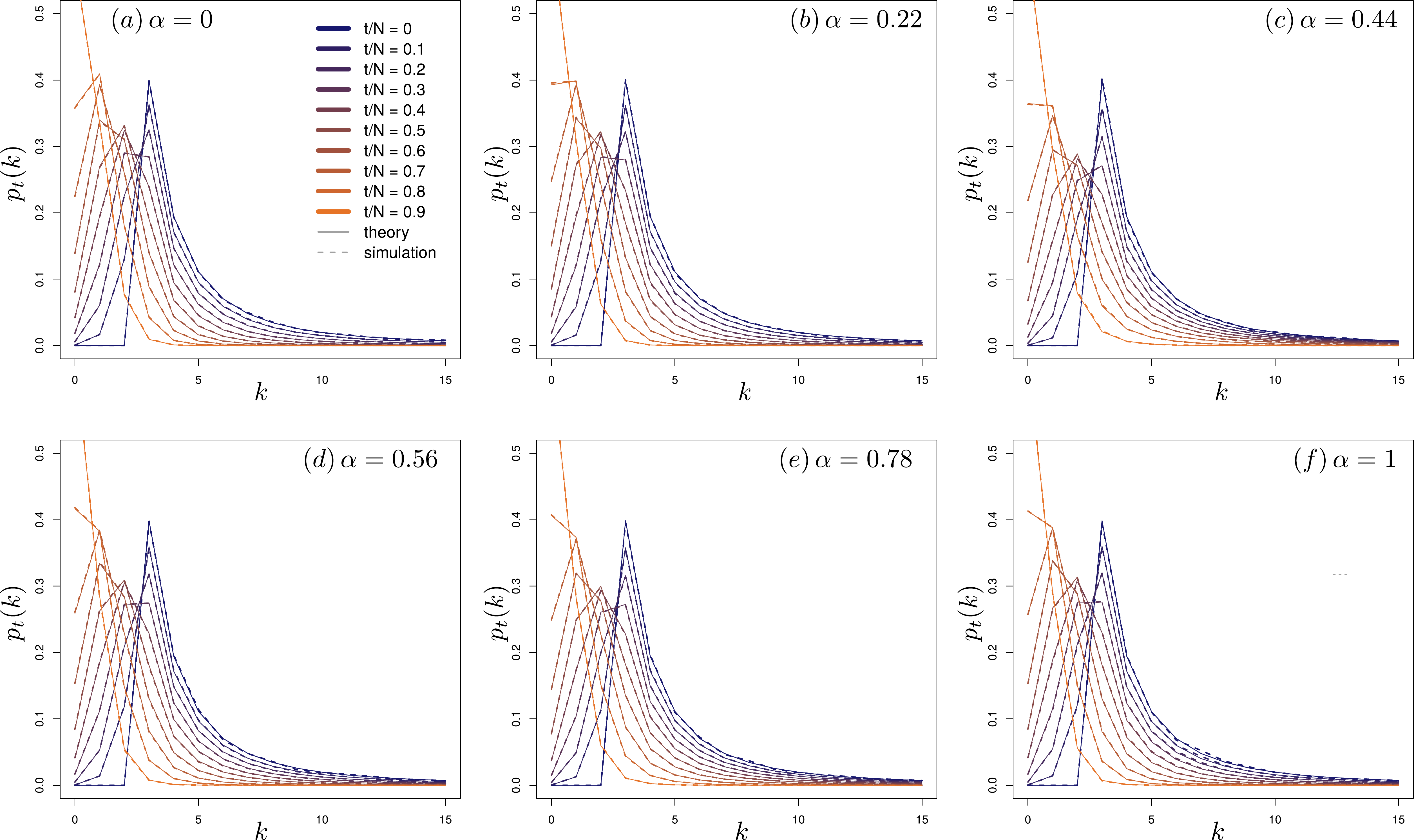}

\end{figure*}

\begin{figure*}
\centering
\captionsetup{justification=raggedright}
\caption{
	\label{fig:SM4}%
	\small
	Comparison between theoretical predictions and empirical estimations of $p_t$ for several $\alpha$ on \textbf{SF} networks with $N_0=10^3$, $\gamma = 2.5$ and $k_{\text{min}} = 3$ in case of \textbf{biased} teleportation.}

    \includegraphics[width=0.9 \linewidth]{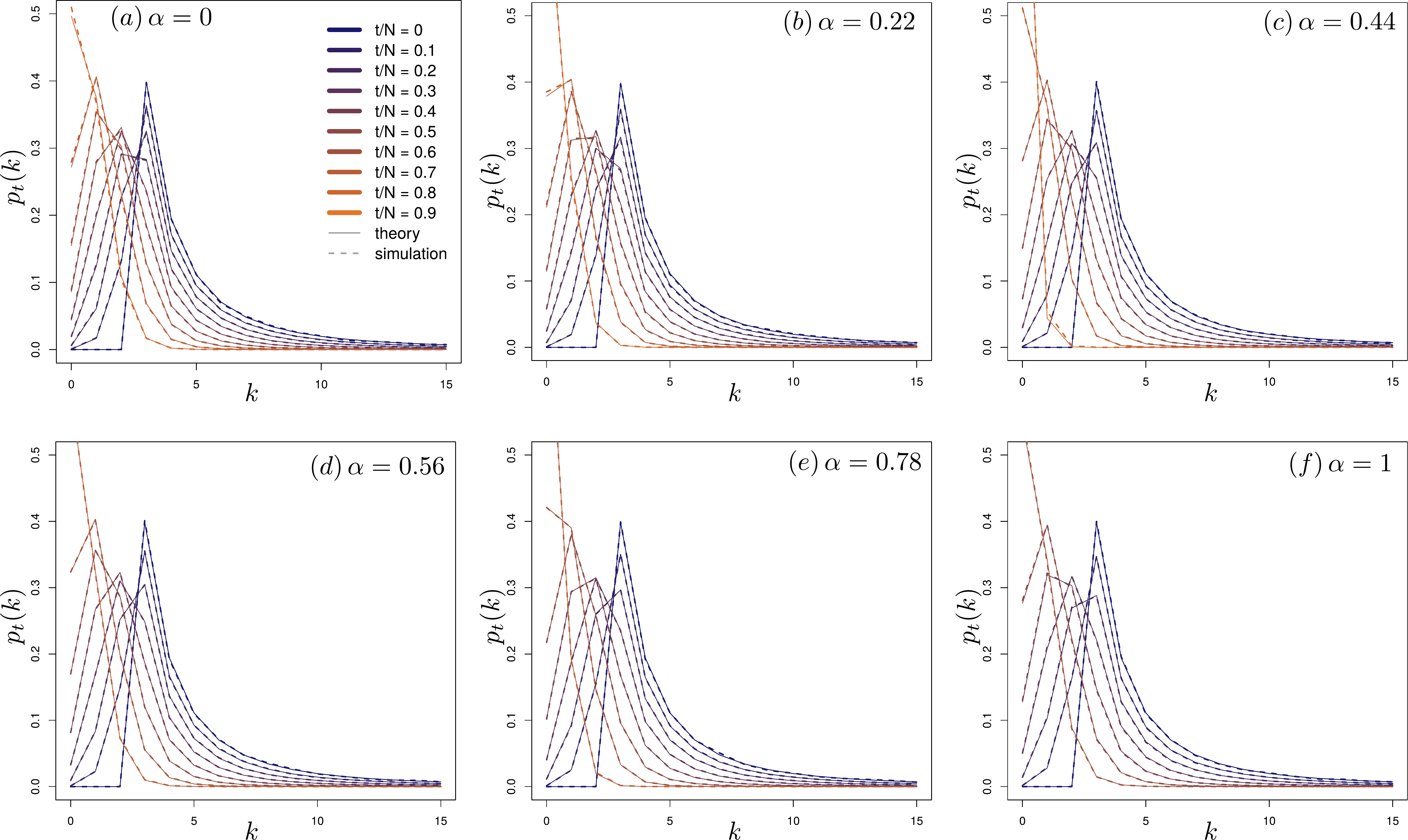}

\end{figure*}

Figs.~\ref{fig:SM1}--\ref{fig:SM4} show the comparison between the theoretical expression of the time-dependent degree distribution $p_t(k)$ and its values when computed directly from the simulations. Several network topologies and teleporting rules are displayed: Erd\H{o}s-R\'enyi (ER) networks with uniform teleportation in Fig.~\ref{fig:SM1} and with biased teleportation in Fig.~\ref{fig:SM2}; scale-free (SF) networks with uniform teleportation in Fig.~\ref{fig:SM3} and with biased teleportation in Fig.~\ref{fig:SM4}. For each case, several cases are presented as a function of the teleportation probability $\alpha$.

\begin{figure*}
	\centering
    \includegraphics[width=0.9 \linewidth]{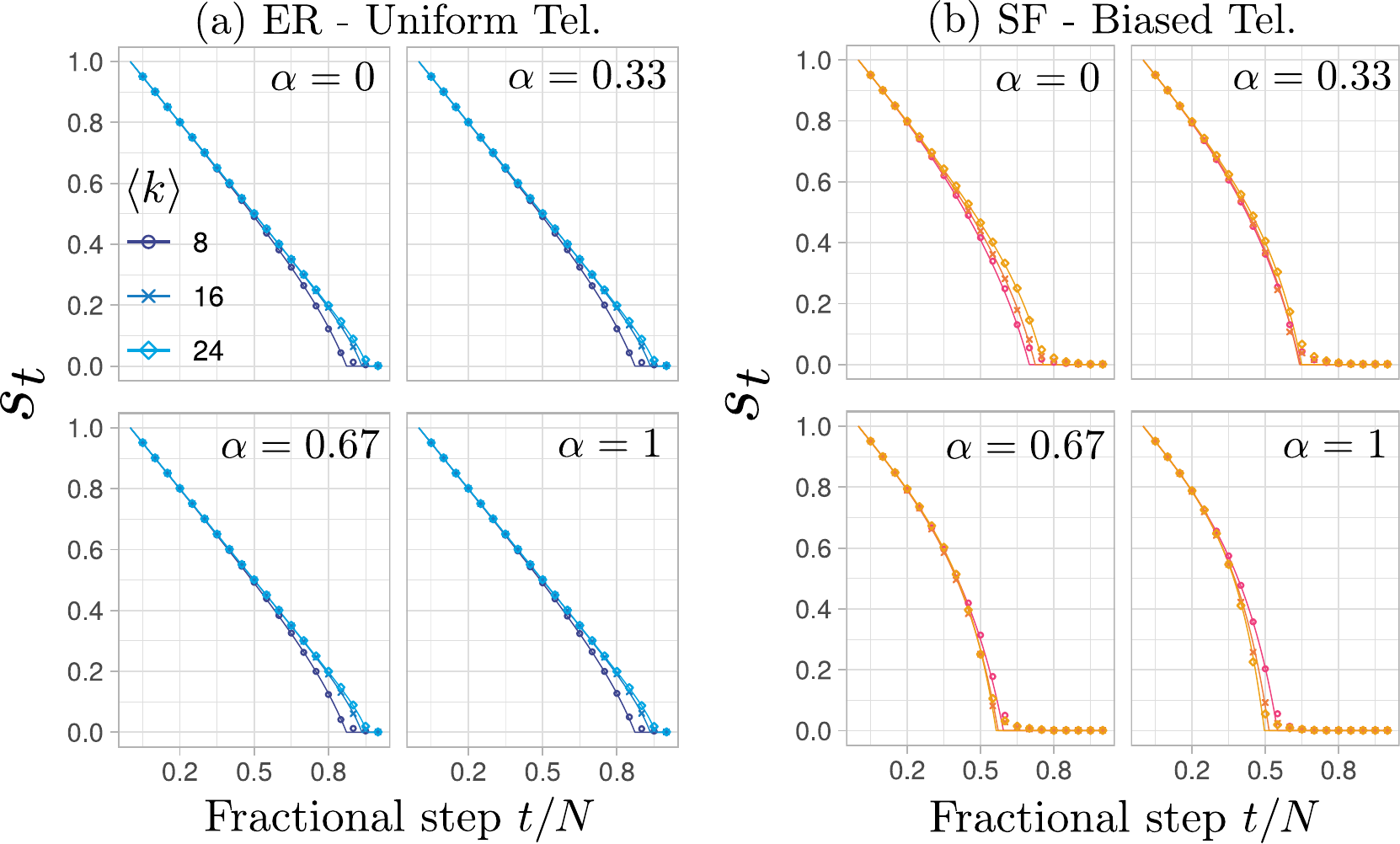}
	\captionsetup{justification=raggedright, singlelinecheck=off}
	\caption{
		\label{fig:SM5}
		Time evolution of the fractional size of the giant component for Erd\H{o}s–Rényi graphs with uniform teleportation (left) and scale-free networks with biased teleportation (right). Different values of the mean degree and the power law exponent are considered, indicated in the legend.
		Initial network size is $N_0=10^3$ nodes, and for the SF networks we consider a minimum degree equal to 3. Solid and dashed lines represent respectively theoretical predictions and simulations.
	}
\end{figure*}

Fig.~\ref{fig:SM5} complements \FigMain{Fig.~2} of the main text, displaying here the two cases not shown there (ER with uniform teleportation in Fig.~\ref{fig:SM5}(a) and SF with biased teleportation in Fig.~\ref{fig:SM5}(b)).

Fig.~\ref{fig:SM6} supports what is discussed in the last section of the Appendix. It shows the stopping time distribution for different network topologies and different teleportating rules.

Finally, Fig.~\ref{fig:SM7} complements \FigMain{Fig.~3} of the main text, providing the single plots for each network topology and teleportation type, and curves with more values of $\langle k \rangle$ and $\gamma$.

\begin{figure*}
	\centering

    \includegraphics[width=0.9 \linewidth]{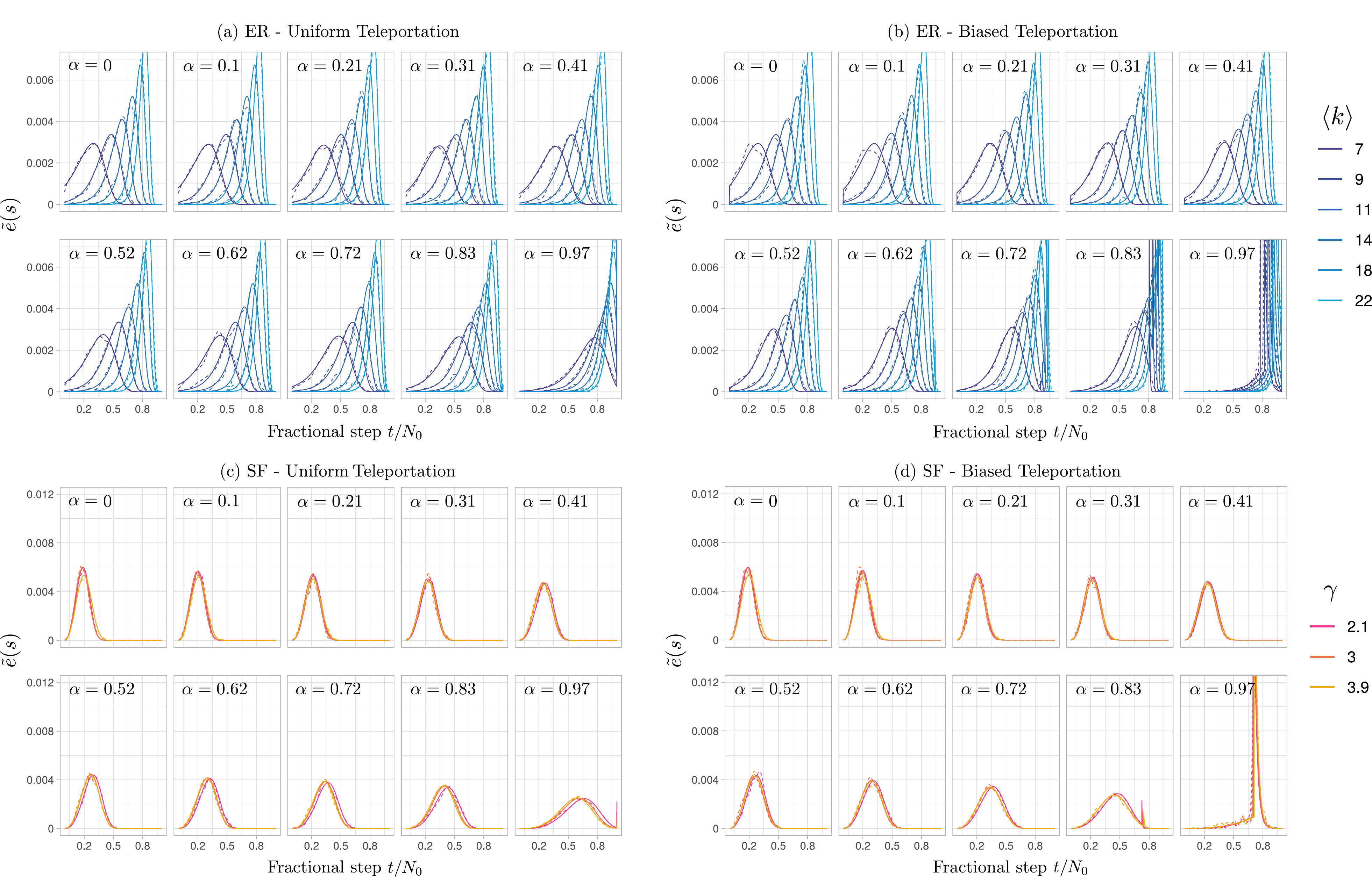}

	\captionsetup{justification=raggedright, singlelinecheck=off}
	\caption{
		\label{fig:SM6}%
		Fractional stopping time distribution $\tilde{e}(s)$ for ER and SF networks with $N_0=10^3$ nodes. Several values of the teleportation parameter $\alpha$, readable at the top of each box, are considered. Solid and dashed lines represent respectively theoretical predictions and empirical estimations.
	}
\end{figure*}

\begin{figure*}
	\centering

    \includegraphics[width=0.9 \linewidth]{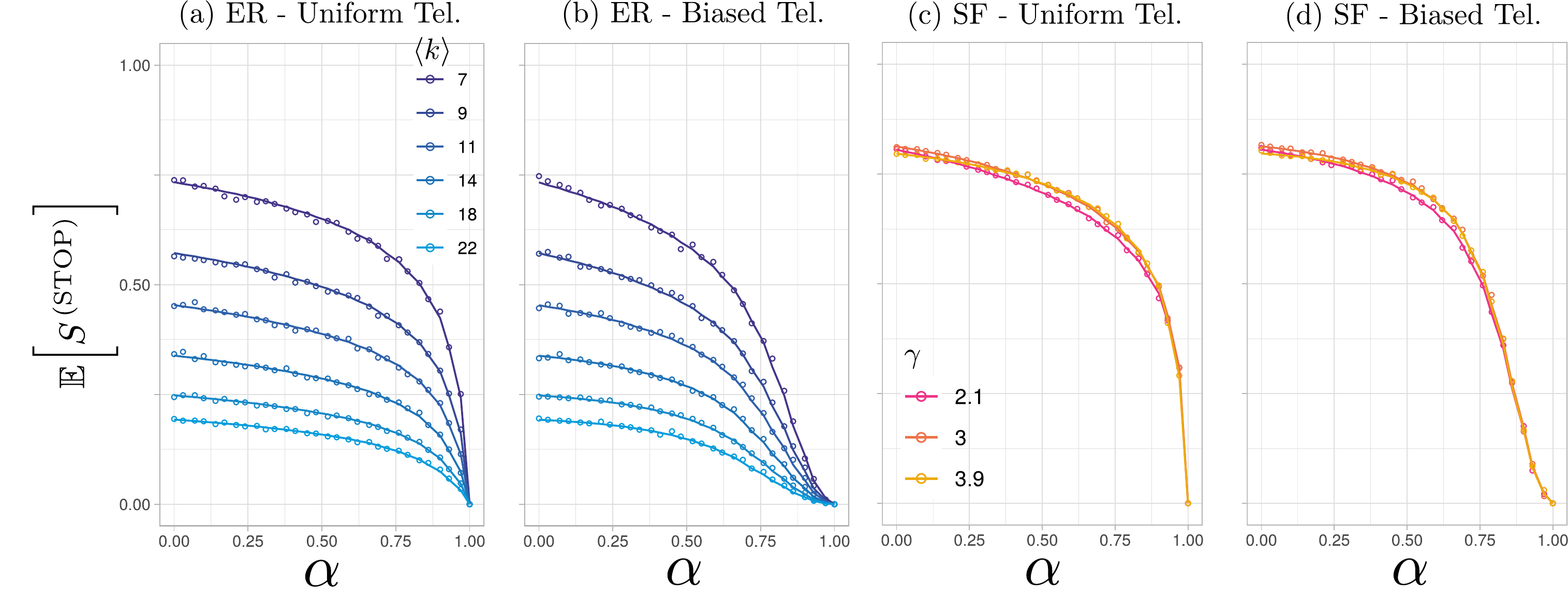}

	\captionsetup{justification=raggedright, singlelinecheck=off}
	\caption{
		\label{fig:SM7}%
		Expected value of size of the giant component at the cascade stop as a function of the teleportation parameter $\alpha$. The network and teleportation type is indicated above each panel. Solid lines come from theory, markers from simulations, averaged over $900$ realizations. Network size is~$N_0=10^3$, and for SF nets we set $k_{\text{min}} = 3$
	}
\end{figure*}

\end{document}